\documentclass[aps,twocolumn,showpacs,floats,longbibliography]{revtex4}
\usepackage{graphicx}
\usepackage{dcolumn}
\usepackage{bm}
\usepackage{color}
\usepackage{amsmath}
\usepackage{amssymb}
\usepackage{hyperref}

\begin{document}

\title{Simulated quantum annealing of double-well and multi-well potentials}

\author{E. M. Inack$^{1,2}$ and S. Pilati$^{1}$}
\affiliation{$^{1}$The Abdus Salam International Centre for Theoretical Physics, I-34151 Trieste, Italy}
\affiliation{$^{2}$SISSA - International School for Advanced Studies and INFN, Sezione di Trieste, I-34136 Trieste, Italy}

\begin{abstract}
We analyze the performance of quantum annealing as a heuristic optimization method to find the absolute minimum of various continuous models, including landscapes with only two wells and also models with many competing minima and with disorder.
The simulations performed using a projective quantum Monte Carlo (QMC) algorithm are compared with those based on the finite-temperature path-integral QMC technique and with classical annealing.
We show that the projective QMC algorithm is more efficient than the finite-temperature QMC technique, and that both are inferior to classical annealing if this is performed with appropriate long-range moves. However, as the difficulty of the optimization problem increases,  classical annealing looses efficiency, while the projective QMC algorithm keeps stable performance and is finally the most effective optimization tool.
We discuss the implications of our results for the outstanding problem of testing the efficiency of adiabatic quantum computers using stochastic simulations performed on classical computers.
\end{abstract}

\pacs{02.70.Uu,02.70.Ss,07.05.Tp,75.10.Nr}
\maketitle


Recent extraordinary developments in the technology of superconducting flux qubits give us well-grounded hope that adiabatic quantum computers capable to solve large-scale optimization problems via quantum annealing will be available in the near future~\cite{johnson2011quantum,boixo2013experimental,boixo2014evidence}.
However, the currently available quantum annealers did not demonstrate superiority with respect to state-of-the-art classical optimization algorithms~\cite{troyerdefining,isakov2015optimised}, and it is still under investigation whether their quantum features play a fundamental functional role in the optimization process~\cite{smolin2013classical,troyerreexamining,wang2013comment,shin2014quantum,lanting2014entanglement}.
In fact, it is not even clear if, at least under certain circumstances, one should expect quantum annealing to be superior to classical methods~\cite{das2008colloquium}, in particular to simulated (classical) annealing~\cite{vecchi}.
Some indications suggesting the supremacy of quantum annealing were originally provided by experiments performed on disordered magnetic materials~\cite{brooke}.
Unfortunately, giving a definite answer to this issue using classical computers is not straightforward~\cite{nishimoriising}, since the computational times required to exactly simulate the real-time dynamics of the quantum annealing process (as defined in the framework of adiabatic quantum computing~\cite{farhi2001quantum}) increase exponentially with the number of variables. Therefore, one has to resort to approximate simulation methods. 
The most relevant one consists in performing stochastic simulations based on quantum Monte Carlo (QMC) algorithms while slowly varying an annealing parameter~\cite{finnila,santorotheory}, thus defining an inhomogeneous Markov chain~\cite{nishimorifoundations}. This mirrors the approach of classical annealing (CA).
As a matter of fact, early QMC simulations of random Ising models based on the path-integral Monte Carlo (PIMC) method with discrete imaginary-times provided results in line with the experiment~\cite{santoroPIMC}.
However, QMC simulations have also provided negative indications. This is the case of the recent PIMC simulations of the random Ising models performed in the continuous imaginary-time limit~\cite{troyerheim}, and of those of the 3-SAT problem~\cite{santoro3sat}. 
Furthermore, it has been argued that in certain conditions path-integral based algorithms might not be able to equilibrate in polynomial times, making the evaluation of the performance of quantum annealing ambiguous~\cite{hastings2013}.\\
It is of outstanding importance to establish if and how computer simulations based on stochastic methods (hereafter referred to as simulated quantum annealing (SQA))  can be used to ascertain the superiority of quantum annealing versus classical algorithms. This would permit us to understand under which conditions quantum speed-up is attainable~\cite{troyerdefining}, and to identify the distinctive signatures of quantum effects to be sought for in a quantum device.\\
In this Article, we tackle these open problems by analyzing the performance of SQA in finding the absolute minimum of simple double-well potentials and in more intricate multi-well potentials with disorder and with competing interaction terms. Compared to the Ising models, such continuous-space potentials allow us to tune more easily 
the difficulty of the optimization problem, an aspect which was indeed found to be of crucial importance for a fair assessment of the performance of quantum annealing~\cite{katzgraberblind}. 
Furthermore, effective double-well and multi-well potentials have recently been implemented on a D-Wave machine~\cite{boixoquantumtunneling}.\\
In most previous studies addressing analogous problems the simulations were performed using the PIMC method (relevant exceptions are Refs.~\cite{finnila,santoroGFMC,sandvikPRL2015}). This is designed to simulate quantum many-body systems at finite temperatures, and is based on an effective classical model which evolves according to the stochastic dynamics defined by the Metropolis algorithm.
Instead, here we employ a projective QMC technique, namely the diffusion Monte Carlo (DMC) algorithm~\cite{kalos}. This is based on the stochastic simulation of the time-dependent Schr\"odinger equation in imaginary time. It permits to sample configurations according to the ground-state wave function, thus providing access to zero-temperature properties.\\
It is worth emphasizing that neither the PIMC nor the DMC algorithms directly simulate the real-time Schr\"odinger dynamics of the quantum annealing process as it would be implemented on an adiabatic quantum computer. In relation to this, they should be regarded as quantum inspired heuristic optimization methods. However, there is a strong connection between the imaginary-time dynamics of the DMC simulations and the real-time dynamics of the adiabatic quantum computation. Indeed, it was shown in Ref.~\cite{nishimorifoundations} that, in the regime of applicability of an adiabatic perturbation theory, the optimization performed via imaginary-time dynamics is (asymptotically) equally effective as the real-time counterpart. It has also been found that certain universal critical properties characterizing the real-time evolution through a quantum critical point can be extracted from the imaginary-time dynamics~\cite{sandvikPRB2011}. Therefore, while the relation between the Metropolis dynamics of the PIMC simulations and the real-time quantum annealing is not evident, the DMC algorithm can provide us with less ambiguous information on the true potential of quantum annealers, at least in the aforementioned circumstances.\\
In tune with this, we compare the performances of the SQA implemented using the DMC and PIMC algorithms (for the latter we employ data from Ref.~\cite{santorodoublewell}) and we highlight their radically different behaviors due to their distinct stochastic dynamics. The DMC quantum annealing is then compared with CA. We analyze how the performances of these two methods degrade when the problem difficulty increases, and we identify some conditions where the DMC algorithm outperforms CA. One of the goals is to single out the features which distinguish quantum annealing from CA.\\
The rest of this Article is organized as follows: in Section~\ref{secmethod} we describe the implementation of SQA with the DMC algorithm, as well as the CA methods we employ to perform comparisons with SQA. 
In Section~\ref{secresults}, we first consider the optimization of both symmetric and asymmetric double-well potentials with different types of SQA and CA methods. 
Then, we address more intricate models with many closely competing minima characterized by an increasing degree of difficulty, including: the multi-well washboard potential, the quasi-periodic (double-sinusoidal) potential, and a two-particle model with competing interaction terms. 
Our conclusions concerning the potential supremacy of quantum annealing and the possibility to analyze its efficiency with stochastic simulations are reported in Section~\ref{secconclusions}.\\

\section{Methods}
\label{secmethod}
The DMC algorithm is one of the most powerful stochastic techniques to simulate the ground state of quantum many-body systems~\cite{anderson,kalos}. It has proven to be extremely effective in numerous studies of divers systems, including electrons in solids, quantum fluids, nuclear matter, ultracold atoms, and also discrete lattice models.\\
In this article we consider one-particle and two-particle continuous-space models in one spatial dimension. The Hamiltonian can be written in the generic form (here, and in the rest of the article, we set $\hbar=1$):
\begin{equation}
\hat{H}=-\frac{1}{2m}\sum_{i=1}^{N}\nabla_i^2+ V({\bf X}),
\end{equation}
where $m$ is the particles mass, ${\bf X}=\left(x_1,\dots,x_N\right)$ denotes the particles configuration, with $x_i$ the position of the particle $i$ (with $i=1,\dots,N$), and $N$ is the particle number. We consider only the two cases $N=1$ and $N=2$.
The total potential-energy operator $V({\bf X})=\sum_{i<j}v_{\mathrm{int}}(\left|{ x}_i-{ x}_j\right|) + \sum_{i}v_{\mathrm{ext}}({ x}_i)$ is composed by the two-body interparticle interaction $v_{\mathrm{int}}(x)$ and by the external potential $v_{\mathrm{ext}}(x)$.\\
The DMC algorithm projects out the ground-state wave function by evolving the following modified time-dependent Schr\"odinger equation for the function $f({\bf X},\tau)=\Psi ({\bf X},\tau)\psi_T({\bf X})$ written in imaginary-time $\tau=i t$:
\begin{eqnarray}
-\frac{\partial f({\bf X},\tau)}{\partial\tau}= &-& D\nabla_{\bf X}^2 f({\bf X},\tau) + D \nabla_{\bf X}[{\bf F}({\bf X})
f({\bf X},\tau)] \nonumber \\
&+& [E_L({\bf X})-E_{\textrm{ref}}]f({\bf X},\tau) \;.
\label{FNDMC}
\end{eqnarray}
Here, $\Psi({\bf X},\tau)$ denotes the wave function at the imaginary time $\tau$ and $\psi_T({\bf X})$ is a trial function used for 
importance sampling. Moreover, $E_L({\bf X})=
\psi_T({\bf X})^{-1}H\psi_T({\bf X})$ denotes the local energy, ${\bf F}({\bf X})=2\psi_T({\bf X})^{-1}\nabla_{\bf X}
\psi_T({\bf X})$ is the quantum drift force, $D=(2m)^{-1}$ plays the role of an effective diffusion constant, while
$E_{\mathrm{ref}}$ is a reference energy. 
The modified Schr\"odinger equation~(\ref{FNDMC}) can be solved by applying iteratively the integral equation:
\begin{equation}
\label{integral}
f({\bf X},\tau+\Delta \tau) = \int \textrm{d} {\bf X}' G({\bf X}',{\bf X},\Delta \tau) f({\bf X}',\tau),
\end{equation}
where $\Delta\tau$ is a short time-step, and $G({\bf X}',{\bf X},\Delta \tau)$ is a suitable approximation (exact in the $\Delta\tau\rightarrow 0$ limit) for the Green's function of Eq.~(\ref{FNDMC}). In this Article, we employ the so-called primitive approximation~\cite{thijssen}:
\begin{equation}
\label{primitive}
G({\bf X}',{\bf X},\Delta \tau) \approx G_d({\bf X}',{\bf X},\Delta \tau)G_b({\bf X}',{\bf X},\Delta \tau)
\end{equation}
where
\begin{multline}
G_d({\bf X}',{\bf X},\Delta \tau) = \\
\left( 4D\pi \Delta\tau \right)^{-N/2} 
\exp \left[ -\frac{\left({\bf X}'-{\bf X}-\Delta\tau {\bf F}({\bf X})/2\right)^2}{4\Delta\tau D}\right]
\end{multline}
and
\begin{equation}
G_b({\bf X},{\bf X}',\Delta \tau)= \exp \left[-\Delta\tau \left(E_L({\bf X}')- E_{\textrm{ref}} \right) \right].
\end{equation} 
Eq.~(\ref{integral}) could be interpreted as the definition of a Markov chain with transition matrix equal to the (positive-definite) Green's function $G({\bf X}',{\bf X},\Delta \tau)$. However, while $G_d({\bf X}',{\bf X},\Delta \tau)$ defines a standard drift-diffusion process, the second term $G_b({\bf X}',{\bf X},\Delta \tau)$ is not normalized. The Markov chain can still be defined in an extended configuration space. One has to evolve a (large) ensemble of copies of the system (typically referred to as random walkers) 
according to the drift-gaussian process, with an additional branching (or killing) process where walker replicas are generated (or annihilated) proportionally to $G_b({\bf X}',{\bf X},\Delta \tau)$. This branching process takes into account the lack of normalization of the Green's function, and causes fluctuations in the random-walker number. For the random-walker branching and for the total population control we follow the standard procedure exhaustively described in Ref.~\cite{thijssen}.\\
After an equilibration time, the walkers sample configurations according to the function $f({\bf X},\tau\rightarrow\infty) = \Psi_0({\bf X})\psi_T({\bf X})$, where $\Psi_0({\bf X})$ is the ground-state wave function.
If $\psi_T({\bf X})$ is chosen to be a good approximation of the ground-state wave function, then the most relevant configurations are sampled more frequently and the fluctuations of the number of random walkers are strongly suppressed, given that $E_L({\bf X})$ is close to a constant. This reduces the computational cost, in particular for large systems.
By tuning $E_{\mathrm{ref}}$, one can adjust the average random-walker number at a desired value $N_{\mathrm{w}}$.
The DMC algorithm can also be implemented without importance sampling by setting $\Psi_T({\bf X}) = 1$, usually at the cost of larger computational times. In this case the modified Schr\"odinger equation~(\ref{FNDMC}) reduces to the standard imaginary-time Schr\"odinger equation.
The potential sources of systematic errors in the DMC algorithm originate from the finite time-step  $\Delta\tau$ and the finite walkers population $N_{\mathrm{w}}$. For all the models considered in this article, we carefully analyzed these effects, and we report data obtained with small enough values of $\Delta\tau$ and large enough values of $N_{\mathrm{w}}\in[10000,20000]$ to be close to the asymptotic exact regime.
An exhaustive description of the DMC algorithm can be found, e.g., in Refs.~\cite{reynolds,kalos,needs,thijssen}, and we refer the interested readers to those reviews for more details.\\

In this article, we are interested in using the DMC algorithm as a heuristic optimization method which searches for the optimal configuration ${\bf X}_{\mathrm{min}}$ where the potential attains its minimum value $V_{\mathrm{min}} = V({\bf X}_{\mathrm{min}})$. This can be achieved by implementing a quantum annealing process, in which quantum fluctuations are gradually suppressed  during the stochastic imaginary-time evolution. 
The suppression of quantum fluctuations can be enforced by reducing the diffusion coefficient $D$, which is equivalent to a particle mass increase. This reduces the quantum delocalization of the particle position, thus favoring random-walkers localization in the configuration ${\bf X}_{\mathrm{min}}$ corresponding to the classical absolute minimum.
$D=D(\tau)$ is now time-dependent in a step-wise manner (the imaginary-time can take only the discrete values $\tau=0,\Delta\tau,2\Delta\tau,\dots$) and in each time-interval the Green's function $G({\bf X},{\bf X}',\Delta \tau)$ corresponding to a time-independent Hamiltonian is employed~\cite{santoroGFMC}. Eq.~(\ref{integral}) now defines an inhomogeneous Markov chain, since the transition matrix varies at each step, due to the (discrete) changes in $D(\tau)$.
Rigorous sufficient conditions for the ergodicity and for the convergence of this quantum annealing method based on the inhomogeneous Markov chain have been derived in Ref.~\cite{nishimoriconvergence}. The corresponding conditions for CA were derived in Ref.~\cite{geman}.\\
In this Article, we implement the following protocol: first, we make the random-walkers population equilibrate by applying the standard DMC algorithm with a constant $D= D_\mathrm{ini}$ for a sufficiently long equilibration time $\tau_{\mathrm{eq}}$, so that the random walkers distribute according to $f({\bf X},\tau_{\mathrm{eq}}) = \Psi_0^\mathrm{ini}({\bf X})\psi_T({\bf X})$, where $\Psi_0^\mathrm{ini}({\bf X})$ is the ground-state wave function at $D=D_\mathrm{ini}$; 
then, we run the DMC algorithm for a (long) annealing time $\tau_{f}$ while decreasing the effective diffusion coefficient after each time-step $\Delta\tau$ according to the step-wise linear law $D(\tau)=D_\mathrm{ini} -\Delta D\tau/\Delta\tau$, where $\Delta D= D_\mathrm{ini}\Delta\tau/\tau_f$. Here, the imaginary time $\tau$ is measured from the end of the equilibration time. At the end of the annealing process the diffusion coefficient vanishes $D(\tau_{f})=0$, while during the last DMC step it is $D(\tau_f-\Delta\tau)=\Delta D$. All quantum annealing simulations reported in Section~\ref{secresults} start with $D_\mathrm{ini}=0.5$ (this is equivalent to an initial mass $m=1$).\\
For an infinitely slow quantum annealing process (corresponding to $\tau_f\rightarrow\infty$), the random-walkers population would follow the adiabatic ground-state wave function at $D(\tau)$ (multiplied times the trial function $\Psi_T({\bf X})$), which gradually shrinks in the minima of the potential landscape; therefore, at the end of the quantum annealing process all random walkers would concentrate in the absolute minimum ${\bf X}_{\mathrm{min}}$~\cite{nishimoriconvergence}. The key issue we investigate is how efficiently the absolute minimum is found for finite $\tau_f$. To quantify the efficiency of the optimization algorithm we measure the average of the potential energies computed in the configurations corresponding to the final random-walkers populations, formally written as: $\bar{V}(\tau_f) = \int \mathrm{d} {\bf X} f({\bf X},\tau_f)V({\bf X}) / \int \mathrm{d} {\bf X} f({\bf X},\tau_f)$.  In the standard DMC formalism this formula would correspond to the mixed estimator of the potential energy. In particular, we analyze the dependence of the residual energy $\epsilon_{\mathrm{res}}=\bar{V}(\tau_f) - V_{\mathrm{min}}$ as a function of the total annealing time $\tau_f$. 
Notice that the total number of DMC steps in the annealing process is $\tau_f/\Delta\tau$ (we use fixed time-steps, in the range $2\Delta\tau D_{\mathrm{ini}}\in[0.01,0.1]$), simply proportional to the total annealing time; this number determines the run time of the simulation on the classical computer.\\

In order to benchmark the performance of the DMC algorithm, we also perform CA simulations. In CA one uses the Metropolis algorithm to sample configurations according to the Boltzmann canonical distribution $P({\bf X}) = \exp\left( -V({\bf X})/T \right)/Z$, where $Z=\int \mathrm{d}{\bf X} \exp\left( -V({\bf X})/T\right)$ and $T$ is the temperature of a fictitious classical statistical system (we chose units such that the Boltzmann constant is $k_B=1$). The temperature is gradually reduced during the simulation, thus removing thermal fluctuations. We adopt a linear annealing schedule of the temperature: $T(\tau) = T_{\mathrm{ini}}\left(1-\tau/\tau_f\right)$, with $\tau=0,1,\dots,\tau_f$ an integer counting the Monte Carlo sweeps (a number of proposed updates equal to the number of variables), and $T_{\mathrm{ini}}$ the initial temperature.
The Markov chain is specified by the transition probability $W\left({\bf X}',{\bf X}\right) = A\left({\bf X}',{\bf X}\right)P\left({\bf X}',{\bf X}\right)$, where $P\left({\bf X}',{\bf X}\right)$ is the probability to propose a move from the configuration ${\bf X}$ to ${\bf X}'$. For a symmetric proposal function, the acceptance probability is $ A\left(  {\bf X}', {\bf X} \right) = \mathrm{min} \left\{ 1, \exp\left[- ( V({\bf X}')-V({\bf X})  )/T\right] \right\}$. We adopt two proposal functions. The first is the box distribution:
\begin{equation}
P\left({\bf X}',{\bf X}\right) = \frac{1}{2\sigma}\Theta(\sigma-\left|x_i'-x\right|) \; ,
\end{equation}
where $\Theta(x)$ is the Heaviside step function; the second is a Lorentzian distribution:
\begin{equation}
P\left({\bf X}',{\bf X}\right) = \frac{1}{\pi}  \frac{ \sigma}{ (x_i'-x_i)^2 + \sigma^2 }.
\end{equation}
The index $i$ labels the particle being (tentatively) displaced from $x_i$ to $x_i'$, and ${\bf X}'=\left(x_1,\dots,x_i',\dots,x_N\right)$. In both cases the parameter $\sigma$ controls the range of the proposed displacements. However, the two distributions determine qualitatively different dynamics, the first one characterized by short-range moves with maximum range $\sigma$, the second one by long-range jumps due to the fat tail of the Lorentzian distribution. Following Ref.~\cite{santorodoublewell}, we vary the range parameter during the annealing process according to the square root law $\sigma(T) = \sigma_{\mathrm{ini}}\sqrt{T/T_{\mathrm{ini}}}$, where $\sigma_{\mathrm{ini}}$ is the initial range parameter. This schedule was found to generate reasonably constant acceptance rates close to the optimal value~\cite{santorodoublewell}. For the box updates, we adopt the initial range parameter $\sigma_{\mathrm{ini}}=2$; for the Lorentzian updates, we use $\sigma_{\mathrm{ini}}=2.9$ for all models considered in this Article, apart for the two-particle model (see Section~\ref{secresults}), for which we use  $\sigma_{\mathrm{ini}}=1.5$.
Analogously to the case of the DMC simulations, we perform the classical annealing using a large ensemble of random walkers. Before starting the annealing process, we let the population evolve according to the Metropolis algorithm at the constant temperature $T_{\mathrm{ini}}=1$, so that the walkers population equilibrates at the Boltzmann thermal distribution. As a measure of the CA efficiency we consider the residual energy $\epsilon_{\mathrm{res}}$, defined (as in DMC quantum annealing) as the average potential energy of the final random-walker population, minus $V_{\mathrm{min}}$. $\tau_f$ is here the number of Metropolis steps in the annealing process. Clearly, in the classical annealing case one could perform serial single-walker simulations.
Both in DMC quantum annealing and in CA we determine the uncertainty on $\epsilon_{\mathrm{res}}$ by repeating a few (typically $5$) simulations starting from different initial random-walker distributions and computing the standard deviation of the (small) population. The resulting error bar is typically smaller than the symbol sizes.\\

One of the key issues we address is whether SQA, which exploits quantum fluctuations to escape local minima, is more or less efficient than CA, which instead exploits thermal fluctuations. In particular, we analyze how rapidly $\epsilon_{\mathrm{res}}$ decreases with the annealing time $\tau_f$. 
One should notice that the annealing times of the quantum and the classical annealing processes cannot be directly compared. Indeed, while in the former case $\tau_f$ is an imaginary-time, in the latter case it is just an integer counting the number of Monte Carlo sweeps during the annealing schedule. Furthermore, depending on the details of the implementation (e.g., serial versus parallel simulations of the random walkers in DMC) the computational times can be different.
However, in general, $\epsilon_{\mathrm{res}}$ decays asymptotically as a power-law of $\tau_f$. Therefore, as in previous works~\cite{santorodoublewell,santorosimplecases}, we will compare the powers characterizing the asymptotic scalings of $\epsilon_{\mathrm{res}}$ in the various annealing algorithms, thus obtaining a measure of their efficiency which is independent of the implementation details and of the scale chosen to measure $\tau_f$.

\section{Results}
\label{secresults}

\begin{figure}
\begin{center}
\includegraphics[width=1.0\columnwidth]{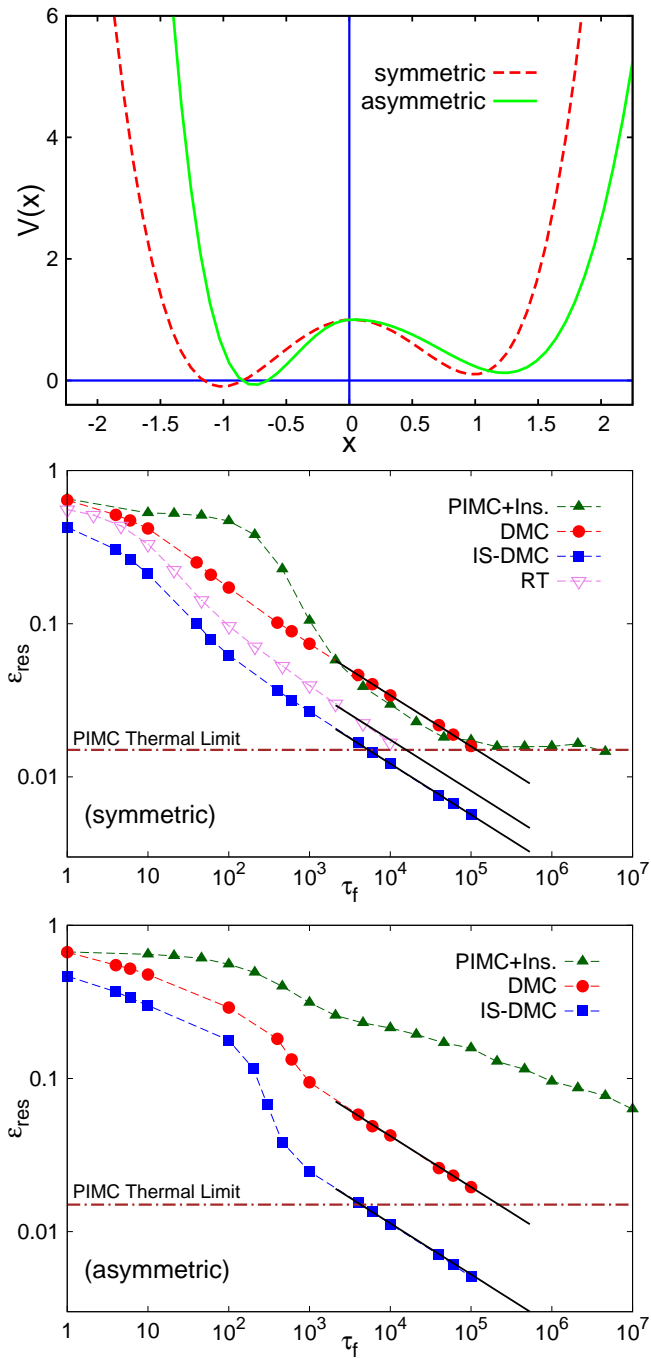}
\caption{(color online). Simulated quantum annealing (SQA) of symmetric and asymmetric double-well potentials.
Top panel: potential energy $V(x)$ versus particle coordinate $x$.
Central panel: residual energy $\epsilon_{\mathrm{res}}$ versus annealing time $\tau_f$ for the symmetric double well, obtained using the diffusion Monte Carlo (DMC) algorithm, the DMC algorithm with importance sampling (IS-DMC), the path-integral Monte Carlo algorithm with instanton move (PIMC+ins., from Ref.~\cite{santorodoublewell}), and via integration of the real-time Schr\"odinger equation (RT, from Ref.~\cite{santorosimplecases}).
Bottom panel: as in the central panel (except for the RT data), for the asymmetric double-well.
The horizontal brown dot-dashed lines indicate the lowest  $\epsilon_{\mathrm{res}}$ reachable in the PIMC simulation due to the finite temperature.
The thick black solid segments indicate fits to the DMC asymptotic data according to the power-law scaling $\epsilon_{\mathrm{res}} \sim \tau_f^{-1/3}$.
The units of $\tau_f$ in PIMC and DMC simulations are different (see text). The thin dashed curves are guides to the eye. Here and in the other graphs the error bars are smaller than the symbol size.
}
\label{fig1}
\end{center}
\end{figure}
%
We start by analyzing the performance of the DMC algorithm as a heuristic optimization method in the context of double-well potentials. We consider the two models introduced in Ref.~\cite{santorosimplecases}. The first is a symmetric double well: $V_{\mathrm{sym}}= V_0\left(x^2-a^2\right)^2/a^4 + \delta x$, where $V_0=1$, $a=1$ and $\delta=0.1$ (see Fig.~\ref{fig1}, top panel). 
Since $\delta a \ll V_0$, the difference in the two minima is $\Delta_V \simeq 2\delta a$, with the absolute minimum located at $x_{\mathrm{min}} \simeq - a$. The two wells have essentially the same widths. 
We adopt the annealing protocol described in the Section~\ref{secmethod}, where the parameter $D=1/(2m)$ is linearly reduced to zero (in a step-wise manner). We perform both DMC simulations with importance sampling (using a Boltzmann-like trial wave function $\Psi_T(x) = \exp\left(-\tilde{\beta} V(x)\right)$, where $\tilde{\beta}=0.8$ is a fictitious inverse temperature) and also without importance sampling, setting $\Psi_T(x) = 1$. 
The results for the residual energy $\epsilon_{\mathrm{res}}$ as a function of the annealing time $\tau_f$ are displayed in Fig.~\ref{fig1} (we recall that the number of DMC steps, and so the simulation run-time, is simply proportional to $\tau_f$). We observe that the use of importance sampling introduces a quantitative improvement, providing somewhat lower residual energies; however, the two approaches (with and without importance sampling) display the same asymptotic scaling $\epsilon_{\mathrm{res}}\propto \tau_f^{-1/3}$, meaning that the efficiency of the optimization process is not affected in a qualitative manner. This power-law dependence with the characteristic power $-1/3$ was first found in Ref.~\cite{santorosimplecases} by exactly solving the imaginary-time Schr\"odinger equation for a single harmonic well; it appears to be a generic feature of DMC quantum annealing in the asymptotic regime.
Fig.~\ref{fig1} also displays the residual energies obtained in Ref.~\cite{santorodoublewell} using the PIMC method, employing a linear annealing protocol as in our DMC simulations. For the PIMC data, the annealing time $\tau_f$ represents the number of Monte Carlo sweeps. One observes that for large $\tau_f$ the PIMC data decay similarly to the DMC results; however, in the $\tau_f\rightarrow \infty$ limit they saturate at the energy corresponding to the temperature at which the PIMC simulations were performed. In principle, this finite temperature could be reduced arbitrarily close to zero (but at the expense of higher computational cost); therefore, we conclude that the DMC and the PIMC quantum annealing methods perform comparably well in this symmetric double-well problem.\\
A more interesting test problem is obtained by introducing an asymmetry in the widths of the two wells; specifically, we consider the asymmetric potential:
\begin{equation}
\label{asypot}
V_{\mathrm{asym}}(x) = 
\left\{
\begin{array}{cc} 
V_0\left(x^2-a_+^2\right)^2/a_+^4 + \delta x &  \mathrm{if}\; x\geqslant 0 \\
V_0\left(x^2-a_-^2\right)^2/a_-^4 + \delta x   &  \mathrm{if}\; x < 0,
\end{array}
\right.
\end{equation}
where the new constants are $a_-=0.75$ and $a_+=1.25$. In this case, the well corresponding to the false minimum is wider than the well corresponding to the absolute minimum (which is located at  $x_{\mathrm{min}} \simeq -a_- - \delta a_-^2/(8V_0)$, to linear order in $\delta$).
In the early state of the annealing process, where zero-point motion is large, the wave function weight is mostly located close to the false minimum, implying a larger probability to find the quantum particle in the wider well. Only in a later stage of the annealing process, where the annealing parameter $D$ is small (corresponding to a large particle mass), the wave function starts concentrating in the deepest (narrower) well. This is reflected in the dependence of $\epsilon_{\mathrm{res}}$ versus $\tau_f$ obtained with the DMC algorithm, which displays different behaviors in the two stages, with the crossover taking place at $\tau_f\approx 10^3$. In the asymptotic regime $\tau_f \rightarrow \infty$, the residual energy decays again with the power-law $\epsilon_{\mathrm{res}}\propto \tau_f^{-1/3}$, just like in the symmetric wells case, both with and without importance sampling.\\ 
It is interesting to observe that the PIMC data display a qualitatively different behavior: the residual energy decreases much slower than in the symmetric wells case, even well before the thermal limit is reached. This occurs in spite of the fact that the PIMC simulations were performed including the so-called instanton Monte Carlo update, which is designed to displace a significant portion of the path-integral across the energy barrier which separates the two wells, exploiting the knowledge of the potential landscape details. In principle, this kind of update should boost the performance of the PIMC simulations, strongly favoring equilibration. However, it is clear that in the framework of the annealing process this is not sufficient; in fact, the transfer to the deepest (but narrower) well is particularly slow.
This dramatic change of efficiency going from symmetric to asymmetric double-wells appears to be a deficiency of the path-integral scheme, rather than a genuine feature of a perfect quantum annealer. Indeed, when quantum annealing is simulated via the DMC algorithm, the asymmetry of the two wells has essentially no effect on the optimization efficiency. In the DMC scheme, the random-walker distribution in the different wells easily equilibrates thanks to the branching process, making it perfectly suited to simulate the optimization of potential energy landscapes in situations where quantum tunneling across energy barriers plays a fundamental computational role. 
Such a case was indeed recently implemented by researchers working with a D-Wave Two chip via an appropriate choice of the couplings between the quantum spins in two unit cells of the Chimera graph~\cite{boixoquantumtunneling}. 
In this experiment, the effective double-well potential varies in time, with the false minimum appearing first, and the absolute minimum appearing at a later time. While classical trajectories would remain trapped in the false minimum, quantum tunneling allows the system to reach the absolute minimum. This setup is slightly different from the double-well model we address here: in our case the potential does not vary with time, but the system is initially attracted towards the well corresponding to the wrong minimum due to its larger width. It is possible that  in the time-dependent potential case considered in Ref.~\cite{boixoquantumtunneling} the PIMC and DMC algorithms would perform equally well.\\
The DMC simulations have a more direct connection with the quantum annealing as understood in the framework of adiabatic quantum computing, which assumes a real-time Schr\"odinger dynamics with a time-dependent Hamiltonian~\cite{farhi2001quantum}. 
Indeed, inspired by the conjecture formulated in Ref.~\cite{santorosimplecases}, Morita and Nishimori~\cite{nishimorifoundations} showed that, from the imaginary-time version of the adiabatic theorem, it follows that the residual energy obtained from the imaginary-time Schr\"odinger dynamics has, for $\tau_f\rightarrow\infty$, the same asymptotic scaling form as the one obtained from the real-time Schr\"odinger dynamics.
Since the DMC algorithm stochastically simulates the imaginary-time Schr\"odinger equation - with the difference that the annealing parameter is decreased in a step-wise manner (see Section~\ref{secmethod}) - it represents a more legitimate benchmark for the performance of the quantum annealing process as it would be implemented on an ideal (perfectly isolated) quantum device operating at zero temperature. 
Clearly, the conditions for the applicability of the adiabatic perturbation theory of Ref.~\cite{nishimorifoundations}, namely that the adiabatic ground-state contribution is the dominant one at all times, might be violated in certain, possibly relevant, cases. However, the strong relation between imaginary-time and real-time dynamics has been, in fact, confirmed in nontrivial simulations of both clean and disordered Ising models driven to critical points~\cite{sandvikPRB2011,sandvikPRL2015}.
As an illustrative example which also confirms this relation, we show in Fig.~\ref{fig1} (central panel) the residual energies for the symmetric double-well case as obtained by performing quantum annealing via real-time Schr\"odinger dynamics. These data were obtained in Ref.~\cite{santorosimplecases} via exact (deterministic) numerical integration of the time-dependent Schr\"odinger equation. As predicted by the theory of Ref.~\cite{nishimorifoundations}, the real-time residual energies share the same asymptotic scaling behavior as the DMC imaginary-time data, the uniform shift being affected by the use of importance sampling and of the mixed energy estimator in the DMC simulations (see Section~\ref{secmethod}).
Therefore, one understands the importance of comparing the performance of the DMC algorithm with the one of CA-based methods in this and in more challenging optimization problems; this will help us understand in which situations quantum annealing has at least the potential to outperform classical algorithms. Also, it is useful to establish if and when the stochastic simulation of the imaginary-time  Schr\"odinger  dynamics on a classical computer via the DMC algorithm becomes unfeasible (e.g., due to the exponential growth on the required random-walker number~\cite{boninsegnimoroni}), because only in such a case one is actually forced to resort to quantum devices.\\
%
%
\begin{figure}
\begin{center}
\includegraphics[width=1.0\columnwidth]{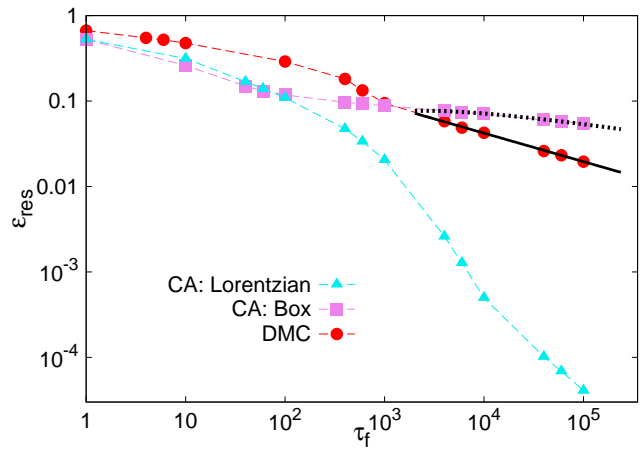}
\caption{(color online). Classical annealing (CA) and SQA of the asymmetric double-well potential.
Residual energy $\epsilon_{\mathrm{res}}$ as a function of annealing time $\tau_f$  obtained using the Metropolis algorithm with Lorentzian proposed moves (cyan triangles) and box-type moves (violet squares), and using the DMC algorithm (red circles).
The thick black solid segments is a fit to the DMC asymptotic data with the power-law $\epsilon_{\mathrm{res}} \sim \tau_f^{-1/3}$.
The dotted black curve is a fit to the CA box-moves data with the asymptotic law $\epsilon_{\mathrm{res}}^{\mathrm{HF}}(\tau_f)$ (see text).
The units of $\tau_f$ in CA and DMC simulations are different, see text.
}
\label{fig2}
\end{center}
\end{figure}
%

Fig.~\ref{fig2} shows the DMC and the CA data for the asymmetric double-well potential. CA is performed both with the short-range (box) and with the long-range (Lorentzian) proposed updates, as described in the Section~\ref{secmethod}. 
In the former case, the stochastic dynamics is well described by the Fokker-Planck equation~\cite{vankampen}, which - as shown by Huse and Fisher~\cite{husefisher} -  in a double-well problem leads to the following asymptotic decay of the residual energy (see dotted curve in Fig.~\ref{fig2}):
\begin{equation}
\label{husefisher}
\epsilon_{\textrm{res}}^{\textrm{HF}}(\tau_f) = c_1\tau_f^{-\Delta_V/B} \left[\ln \left(c_2\tau_f\right)\right]^{2\Delta_V/B},
\end{equation}
where $c_1$ and $c_2$ are fitting parameters, $\Delta_V = \delta (a_+ + a_-)$ is the splitting between the two minima and $B=V_0-\Delta_V-V(x_{\mathrm{min}})$ is the energy barrier separating them.
As discussed above, DMC quantum annealing displays the power-law asymptotic decay  $\epsilon_{\mathrm{res}} \sim \tau_f^{-1/3}$, clearly outperforming CA with short-range updates. However, long-range (Lorentzian) updates strongly increase the efficiency of CA, making it more performant than the DMC algorithm.\\

%
\begin{figure}
\begin{center}
\includegraphics[width=1.0\columnwidth]{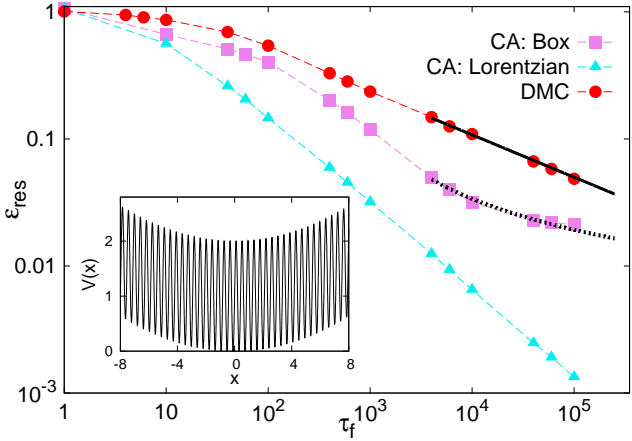}
\caption{(color online). CA and SQA of the washboard potential (shown in the inset).
The symbols are defined as in Fig.~\ref{fig2}.
The thick black solid segment is a fit to the DMC asymptotic data with the power-law $\epsilon_{\mathrm{res}} \sim \tau_f^{-1/3}$.
The dotted black curve is a fit to the CA data with the logarithmic law $\epsilon_{\mathrm{res}} =c_1 \log^{-1}(c_2\tau_f)$, where the fitting parameters are $c_1$ and $c_2$.
}
\label{fig3}
\end{center}
\end{figure}
It is now natural to wonder how CA and DMC quantum annealing perform in more challenging optimization problems. To address this question, we consider the multi-well problem defined by the following ``washboard'' potential (shown in the inset of Fig.~\ref{fig3}):
\begin{equation}
\label{washboard}
V(x) = a_1 x^2 + a_2 \sin\left(a_0 x\right) + a_2,
\end{equation}
where $a_0 = 15$, $a_1 = 0.01$ and $a_2 = 1$. This model was first studied in Refs.~\cite{kabashima1,kabashima2}, where it was suggested that CA should suffer from a pathological slowdown due to the presence of many well-separated minima. In a wide range of annealing times~\cite{santorosimplecases}, the residual energy decay should be at best logarithmic: $\epsilon_{\mathrm{res}}(\tau_f) \propto (\ln(\tau_f))^{-1}$. The CA data with short-range (box) updates (shown in Fig.~\ref{fig3}) are indeed consistent with this logarithmic upper bound, showing that, in general, with the CA dynamics it becomes problematic to equilibrate to the minimum energy configuration when many close solutions compete. Instead, DMC quantum annealing maintains its efficiency, displaying again the asymptotic decay  $\epsilon_{\mathrm{res}} \sim \tau_f^{-1/3}$ (we only display data obtained with the pure DMC algorithm, since importance sampling was again found not to affect the asymptotic efficiency). This $-1/3$ power appears to be the footprint of quantum annealing. Below, we will demonstrate the same behavior in even more intricate problems. 
This suggests that the identification of a $-1/3$ power-law decay in a quantum annealer could be interpreted as an evidence of quantum effects playing a fundamental computational role.
We also notice that, as in the double-well case, long-range updates boost the efficiency of CA at the point of outperforming the DMC algorithm. One might suspect that long-range updates do not provide the same boost in more complex problems with more variables. We will show later on that this is indeed the case.\\

%
\begin{figure}
\begin{center}
\includegraphics[width=1.0\columnwidth]{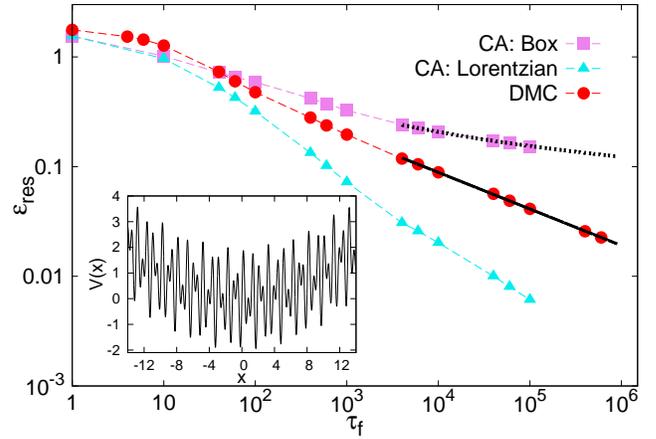}
\caption{(color online). CA and SQA of the quasi-disordered potential (shown in the inset).
The symbols and the curves are defined as in Fig.~\ref{fig2}.
}
\label{fig4}
\end{center}
\end{figure}
%
%
DMC quantum annealing has so far displayed a surprisingly stable efficiency. Clearly, the models we addressed previously do not contain one of the most relevant ingredients which make realistic optimizations problems difficult, namely a disordered distribution of the competing solutions. 
Disorder is indeed expected to hamper quantum annealing due to the Anderson localization phenomenon. This consists in the spatial localization of the Hamiltonian eigenstates, causing the absence of particle diffusion. While in three and more dimensions Anderson localization takes place only for sufficiently strong disorder~\cite{abrahams}, in one and two dimensions any amount of uncorrelated disorder induces localization.\\
A minimal model which contains (quasi) disorder is the following double-sine potential:
\begin{equation}
V(x) = K_0 x^2 +  A\left[  \sin(b_1\pi x) + \sin(b_2 \pi x) \right],
\label{doublesine}
\end{equation}
where $K_0=0.01$ , $A=1$, $b_1=2$,  and $b_2=1+5^{1/2}$. Due to the irrational ratio of the wave-lengths of the two sinusoidal functions, this potential is aperiodic. However, it is deterministic and, therefore, not truly random. This kind of pseudo-randomness is conventionally referred to as quasi-disorder. In a tight binding scheme (which would be rigorously justified if one sinusoidal potential was much more intense than the other one) this incommensurate double-sine model could be approximated by the so-called Aubry-Andr\'e Hamiltonian, provided one neglects the weak harmonic confinement. Differently from one-dimensional models with uncorrelated disorder, in the Aubry-Andr\'e Hamiltonian, Anderson localization takes place at a finite disorder strength or, equivalently, when the particle mass exceeds a critical value~\cite{aubry}. 
We expect the weak harmonic term to play a minor role, at least well beyond the critical point, meaning that the Hamiltonian eigenstates would still be strongly localized in this regime. This strong spatial localization inhibits diffusion, preventing the particle from exploring the complete configuration space, possibly causing localization in local minima.
It is therefore interesting to analyze whether in the last part of the quantum annealing process, where the annealing parameter is small and, correspondingly, the particle mass is large, Anderson localization deteriorates the efficiency of DMC quantum annealing.
%
\begin{figure}
\begin{center}
\includegraphics[width=1.0\columnwidth]{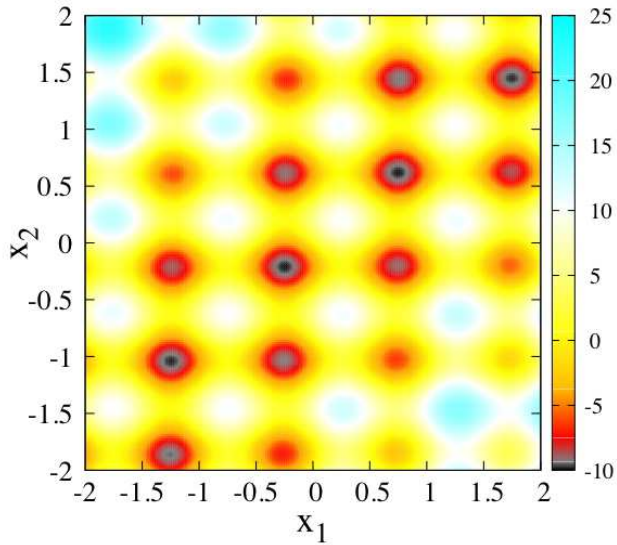}
\caption{(color online). Potential energy $V(x_1,x_2)$ of the two particle model~\ref{twoparticle} in the case $A = 5$. The colorscale represents the potential intensity, $x_1$ and $x_2$ are the coordinates of the two particles.
The absolute minimum is at $V(-0.25,-0.208) \cong -9.99759$.
}
\label{fig5}
\end{center}
\end{figure}
%
%
In fact, the data displayed in Fig.~\ref{fig4} show that the efficiency of DMC quantum annealing is not affected by the presence of quasi-disorder, demonstrating again the extreme stability of its performance (notice that here and in the following we only consider data obtained without importance sampling). As in the case of ordered minima (i.e., the washboard potential) CA with short-range updates displays a pathological slow-down of the annealing process at large annealing times, leading to a logarithmic decay of the residual energy. However, CA with long-range updates is still the most efficient optimization algorithm.\\
%
\begin{figure}
\begin{center}
\includegraphics[width=1.0\columnwidth]{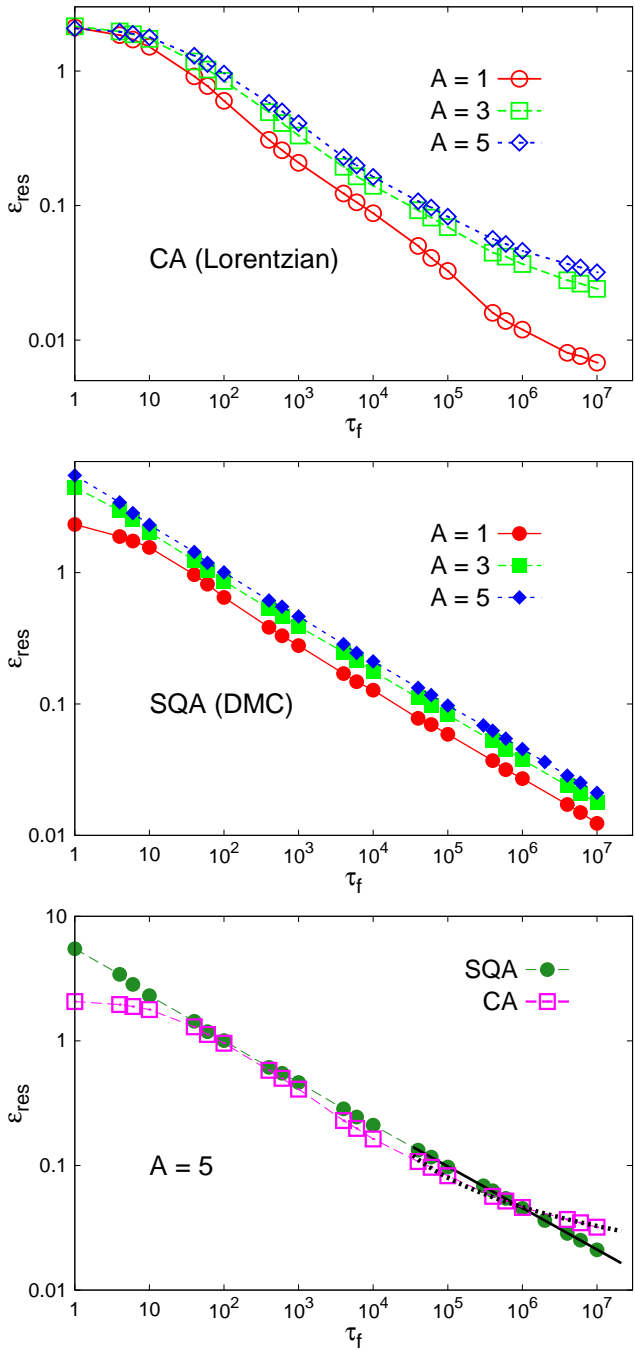}
\caption{(color online). CA and SQA of the two particle model. 
Top panel: residual energy $\epsilon_{\mathrm{res}}$ versus annealing time $\tau_f$ obtained using CA with Lorentzian updates, for different intensities $A$ of the sinusoidal part of the potential.
Central panel: data analogous to those in the top panel, obtained using the DMC algorithm.
Bottom panel: comparison between CA and SQA at $A = 5$. The thick black solid segment indicates the asymptotic scaling of the DMC data $\epsilon_{\mathrm{res}}\sim \tau_f^{-1/3}$, while the black dotted curve the one of the CA data $\epsilon_{\mathrm{res}}\sim \log^{-1} \left(c\tau_f\right)$.
}
\label{fig6}
\end{center}
\end{figure}
\begin{figure}
\begin{center}
\includegraphics[width=1.0\columnwidth]{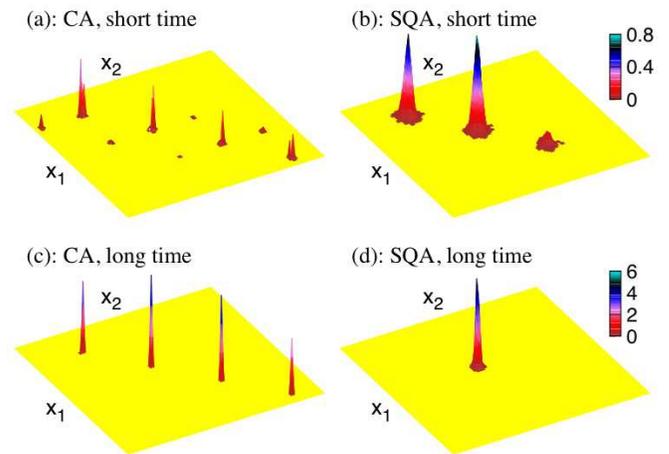}
\caption{(color online). Final random-walker distribution in CA [panels (a) and (c)] and in SQA [panels (b) and (d)]. Panels (a) and (c) correspond to short annealing times $\tau_f = 10^3$, while panels (b) and (d) correspond to long annealing times $\tau_f = 10^5$. The colorscale corresponds to the random-walker density.
}
\label{fig7}
\end{center}
\end{figure}

So far, the DMC algorithm has proven to be an effective and stable optimization method. However, the fact that it is outperformed by CA if appropriate (long-range) Monte Carlo updates are included is quite discouraging for the prospect of developing quantum enhanced optimization algorithms. As anticipated above, one might wonder whether it is always possible to obtain such a boost in the efficiency of CA via long-range updates. To address this question, we consider a model with two particles which move in an external potential and interact with each other. The potential energy is defined as:
\begin{equation}\label{twoparticle}
\begin{split}
V(x_1, x_2) = & K_0(x_1^2 + x_2^2)/2 + K_{\mathrm{rel}} (x_1-x_2)^2/2 + \\
& A \left[\sin ( b_1 \pi x_1) + \sin(b_2 \pi x_2) \right],
\end{split}
\end{equation}
where we set $K_0 = 0.01$, $K_{\mathrm{rel}}= 2$, $b_1=2$, and $b_2=1+2^{1/2}$. For the intensity parameter $A$, we will employ the three values $A = 1,\; 3,\;5$. The first term in Eq.~(\ref{twoparticle}) confines both particles in a global harmonic trap. The second term introduces an attractive harmonic interaction between the two particles. The last term is inspired by the incommensurate double-sine potential analyzed previously. However, here the two particles experience the two sinusoidal fields separately. The first sinusoidal field acts only on one particle, while the second one acts on the other particle. While the attractive interaction tends to localize the two particles in the same location, the sinusoidal terms have their minima in different points. The competition among the terms in Eq.~(\ref{twoparticle}) constitutes the minimal element of frustration. If one interprets  Eq.~(~\ref{twoparticle}) as the external potential of one particle moving in a two-dimensional system, one obtains the intricate landscape shown in Fig.~\ref{fig5}. One notices that there are several closely competing solutions, well separated by high energy barriers. By construction, there is no periodicity and the bottoms of the different valleys are at close but different levels. The height of the barriers can be varied by changing the intensity parameter $A$, allowing us to tune the difficulty of the optimization problem. 
The results of the optimization of this two-particle model are displayed in Fig.~\ref{fig6}, both for CA (top panel) and SQA performed via the DMC algorithm (central panel).
The CA simulations are performed with Lorentzian updates. We apply three kinds of updates: the first displaces only the first particle, the second the other particle, the third applies the same displacement to both particles.
It is evident that when $A$ increases the asymptotic slope of the CA data diminishes, indicating a loss of efficiency of the optimization process. Instead, the DMC data display the same asymptotic decay $\epsilon_{\mathrm{res}} \sim \tau_f^{-1/3}$ for all values of $A$. As anticipated before, the independence of the asymptotic power-law decay of the residual energy appears to be the hallmark of quantum annealing.
The comparison shown in the bottom panel of Fig.~\ref{fig6} (for $A=5$) demonstrates that in the most challenging optimization problem DMC quantum annealing outperforms CA, suggesting that quantum annealing has indeed the potential to outperform classical algorithms in hard optimization problems.
It is particularly instructive to analyze how and why SQA outperforms CA. In Fig.~\ref{fig7} we show the probability to reach a certain two-particle configuration $(x_1,x_2)$, both with CA and SQA, after a short and after a long annealing time. In the mixed estimator scheme (see Section~\ref{secmethod}), this probability corresponds to the spatial distribution of the random-walker population at the end of the annealing process. We observe that, in CA, several false minima have a large probability to be selected. While the random-walker distribution in each individual well rapidly shrinks, even after a long annealing time quite a few competing solutions are still likely to be chosen. This indicates that CA has high chance to remain trapped close to false minima. The behavior of DMC quantum annealing is, in a sense, the opposite: already after a short annealing time only three wells are populated by random walkers. However, the distribution in each well is quite broad, indicating that SQA is slower in sinking to the very bottom of the well. After a long annealing time, only the well corresponding to the absolute minimum is populated, but the residual energy is still nonzero since the random walkers need further time to sink to the very bottom of the well, thus selecting the optimal configuration.\\
At this point, it is worth mentioning  that other classical algorithms which potentially outperform CA for specific optimization problems, in particular for small systems, do exist (e.g., genetic algorithms). However, CA has proven to be one of the most powerful and versatile optimization methods~\cite{vecchi}, succeeding even in challenging continuous-variables problems with multiple minima  (e.g.,  the optimization of the structure of Lennard-Jones clusters~\cite{Wille1987405}) where gradient-based algorithms like the conjugate gradient method remain trapped in local minima. Therefore, CA represents a fair term of comparison for SQA.\\
%
%

\section{Conclusions}
\label{secconclusions}

We have analyzed the efficiency of SQA in finding the absolute minimum of different model potentials in continuous space, including symmetric and asymmetric double wells, and also more intricate models with many closely competing minima with both ordered and disordered spatial distribution of the wells. Contrarily to the finite-temperature path-integral Monte Carlo techniques adopted in several previous studies, the projective method employed in this work, namely the DMC algorithm, exhibits a stable performance which is not affected by details of the potential energy landscape like the asymmetry of the competing wells. Furthermore, due to the formal relation between imaginary-time and real-time Schr\"odinger dynamics (valid in quasi-adiabatic regimes~\cite{nishimorifoundations}), the outcomes of the DMC simulations (which are based on the imaginary-time dynamics) have more direct implications for the evaluation of the potential of adiabatic quantum computing.\\
While the DMC quantum annealing is outperformed by CA in simple one-variable model potentials if one employs ad-hoc Monte Carlo updates which exploit the specific features of the potential landscapes, it is easy to construct more challenging optimization problems with more variables where such tricks in CA become ineffective and SQA turns out to be the most effective optimization method. This result is strongly encouraging for the prospect of developing quantum devices which exploits quantum fluctuations to enhance the efficiency of optimization methods. 
The stable performance of DMC quantum annealing, characterized in quite general continuous-space models by an asymptotic power-law decay of
the residual energy, appears to be a hallmark of an ideal quantum annealer. 
The DMC algorithm is designed to simulate the ground states of isolated quantum systems; it does not take into account thermal fluctuations nor the coupling to the environment. Including these effects in a projective QMC algorithm would provide us with an extremely useful tool to investigate the potential of realistic devices designed to perform adiabatic quantum computations. This is clearly an interesting direction for future research. 
Also, it would be important to identify the cases where the DMC simulations become infeasible due to, e.g., an exponential scaling of the required random-walker number~\cite{nemec}. Computational problems of this kind did not occur in any of the double-well and multi-well problems addressed in this work, probably due to the small particle number.
However, in the context of standard (i.e., without annealing) quantum simulations of parahydrogen clusters, results suggesting an exponential scaling of the random-walker population with the system size have been reported~\cite{boninsegnimoroni}. Also, frustration is expected to harm the efficiency of the DMC simulations. 
Further exploring such potential pathologies of the DMC algorithm in the context of quantum annealing simulations could shed light on the features which make a problem hard for optimization methods based on quantum fluctuations, and also on the conditions where adiabatic quantum computers are expected to outperform stochastic simulations performed on classical computers.\\

We acknowledge insightful discussions with G. Mossi, G. Santoro, A. Scardicchio, M. Troyer.


\end{document}